\documentclass[12pt]{iopart}

\usepackage{amssymb}
\usepackage{color}
\usepackage{epsf}
\usepackage{epsfig}
\usepackage{graphicx}
\usepackage{tikz}

\renewcommand{\d}[1]{\mathinner{d#1}}
\newcommand{\fn}[2]{\mathinner{#1\mathopen{\left(#2\right)}}}

\newcommand{\eq}[1]{Eq.~(\ref{#1})}
\newcommand{\eqs}[2]{Eqs.~(\ref{#1}) and (\ref{#2})}
\newcommand{\eqss}[3]{Eqs.~(\ref{#1}), (\ref{#2}) and (\ref{#3})}
\newcommand{\bvect}[1]{{\hat{\bf{k}}_{#1}}}
\newcommand{\bvectx}[1]{{\hat{\bf{x}}_{#1}}}

\newcommand{\eV}{\mathinner{\mathrm{eV}}}

\newcommand{\GeV}{\mathinner{\mathrm{GeV}}}
\newcommand{\TeV}{\mathinner{\mathrm{TeV}}}
\newcommand{\Hz}{\mathinner{\mathrm{Hz}}}
\newcommand{\km}{\mathinner{\mathrm{km}}}

\newcommand{\myfigure}[2]{\resizebox{#1}{!}{\includegraphics{#2}}}

\begin{document}

\title{Thermal~Inflation~and~the Gravitational~Wave~Background}

\author{Richard Easther$^1$, John T. Giblin, Jr$^1$, Eugene A. Lim$^2$,\\ Wan-Il Park$^3$ and Ewan D. Stewart$^3$}

\address{$^1$ Department of Physics, Yale University, New Haven, CT 06520, USA}
\address{$^2$ ISCAP and Physics Department, Columbia University, New York, 10027 NY.}
\address{$^3$ Department of Physics, KAIST, Daejeon, South Korea}

\begin{abstract}
We consider the impact of thermal inflation -- a short, secondary period of inflation that can arise in supersymmetric scenarios -- on the stochastic gravitational wave background.    We show that while the primordial inflationary gravitational wave background is essentially unchanged at CMB scales, it is massively diluted at solar system scales and would be unobservable by a BBO style experiment. Conversely, bubble collisions at the end of thermal inflation can generate a new stochastic background. We calculate the likely properties of the bubbles created during this phase transition, and show that the expected amplitude and frequency of this signal would fall within the BBO range.
 \end{abstract}

\maketitle

\section{Introduction}

In the coming decades, it is very likely that gravitational waves will be observed by direct-detection experiments. These include the presently operating LIGO and VIRGO observatories, and proposed experiments such as LISA \cite{LISA}, BBO \cite{BBO} and DECIGO \cite{Kawamura:2006up}. It is likely that the first sources to be detected will be localized, transient events, but many cosmological processes yield a stochastic background of gravitational waves. These backgrounds are a key target for the nascent field of gravitational astronomy, since they probe the primordial universe in  ways that are presently impossible.

The most widely discussed cosmological signal is the inflationary gravitational wave background, generated by quantum fluctuations of spacetime. The amplitude of this signal decreases with the inflationary energy scale, rendering it effectively invisible in models where inflation occurs well below the GUT scale.  A stochastic background of gravitational waves can also be generated whenever the universe undergoes a phase transition. Unlike the inflationary background, which is scale invariant, these backgrounds will be peaked at a wavelength correlated with the characteristic scale of the transition. In these scenarios, the gravitational waves are sourced  by the non-zero quadrupole (and higher) moments of the matter distribution \cite{WeinbergGRBook}.

Recently, several groups have investigated the possibility that gravitational waves are produced during the preheating phase at the end of inflation \cite{GarciaBellido:2007af,Easther:2006gt,Easther:2006vd,Khlebnikov:1997di,Dufaux:2007pt,
GarciaBellido:2007dg,Easther:2007vj}.  The momentum modes of fields coupled to the inflaton undergo resonant pumping, which renders the universe highly inhomogeneous, and sources the emission of gravitational radiation.   Intriguingly, this signal is  {\em easier\/} to detect in models where the inflationary scale is substantially below the GUT scale, as first argued in \cite{Easther:2006gt}, and confirmed in \cite{Easther:2006vd}.   Moreover, it probes the mechanism that \emph{ends\/} inflation, which is normally assumed to be inaccessible to direct observational tests.

In addition, a second (and not entirely unrelated) mechanism could generate a significant background -- namely the collision of percolating bubbles formed during a first order phase transition   \cite{Kosowsky:1991ua,Turner:1992tz,Kosowsky:1992rz,Kosowsky:1992vn,
Kamionkowski:1993fg,Caprini:2007xq}. As we will see, it has the same scaling properties as the preheating signal -- namely the amplitude need not depend on the energy scale, but its (present-day) wavelength decreases linearly with this scale. Assuming an energy scale somewhere between the electroweak and GUT scales, the Hubble scale will lie between $10^{-5}\eV$ and $10^{14}\GeV$. We will see that if the characteristic bubble radius is within a few orders of magnitude of the Hubble scale at which they form,  the resulting gravitational wave background will have a present-day frequency in the range $10^{-3}$ to $10^{9} \Hz$ \cite{Easther:2006gt,Kamionkowski:1993fg}.

To predict the primordial background, one must know the transfer function, which is determined by the effective equation of state of the universe between the epoch when the signal is generated and the present day. A further complication is that there is no guarantee that inflation happens only once in  the early universe. One may imagine primary and secondary periods of inflation -- where the primary period is responsible for solving the usual cosmological initial conditions problems and laying down the perturbations that seed large scale structure, and the secondary period lasts for $\leq \mathcal{O}(10)$ $e$-foldings, and occurs at some energy scale below that of the first inflationary period and above (or at) the scale at which baryogenesis takes place.  If significantly more than  $\mathcal{O}(10)$ $e$-folds occurs in the secondary phase, it will erase all relics of the previous phase, returning us to what is effectively a single phase model.  In addition to modifying the mapping between scales in the present epoch and the moment these (comoving) scales leave the horizon during the primary phase, a secondary phase of inflation also erases the primordial inflationary background of gravitational waves at wavelengths which have reentered the horizon before the secondary phase begins.  This leads to a situation where a high scale primary phase would generate a gravitational wave background visible in the B-mode of the CMB polarization, but completely \emph{invisible\/} to direct detection experiments operating at solar system scales, such as BBO.

A cosmological model with multiple inflationary phases may appear  baroque. However \emph{thermal inflation} \cite{Lyth:1995hj,Lyth:1995ka} is a secondary period of inflation that arises in  supersymmetric theories when some (almost) flat direction with negative mass-squared at the origin -- which we will call the flaton --  acquires a thermal contribution to its effective potential, giving it a positive effective mass.   The flaton is thus pinned at the origin, and the finite potential energy at this point drives a secondary period of inflation. Once thermal inflation sets in, the temperature falls rapidly and when it drops below the flaton's mass scale (usually assumed to be a typical soft supersymmetry breaking mass scale $\sim \TeV$) the flaton rolls aways from the origin, and thermal inflation naturally shuts off.   The total amount of inflation is typically small, about 10 $e$-folds or less.

Thermal inflation requires only a flat direction with negative mass-squared at the origin, both of which are natural in supersymmetric models.  From the model building perspective, its importance lies in its ability to dilute the moduli and gravitinos, which are generically produced after the primary period of inflation in supersymmetric theories. These particles either decay, wrecking the successful predictions of Big Bang Nucleosynthesis, or provide a dark matter density that massively overcloses the present universe \cite{Coughlan:1983ci,Banks:1993en,de Carlos:1993jw,Khlopov:1984pf,Ellis:1984eq}.  These particles are produced before thermal inflation begins and are too weakly coupled to be produced after it ends, so their contribution is thus diluted to safe levels.  Thermal inflation is perhaps the best-motivated solution to the moduli problem, which is endemic to supersymmetric models of the early universe, including many of those derived from string theory. Thermal inflation has a relatively low reheating scale (typically $\mathcal{O}(10) \GeV$) but it can give rise to baryogenesis \cite{Jeong:2004hy,Felder:2007iz}.

Unfortunately, as we will show in Section~\ref{sect:forecasts}, thermal inflation wipes out any primordial inflationary gravitational wave signal in the range of frequencies accessible to BBO \footnote{Indeed it also wipes out the primordial \emph{density} spectrum at the same frequencies, although the scales affected are not important cosmologically and hence not constrained by large scale structure observations.}. Thus, even if the primary phase of inflation occurs at a  high enough scale to produce a detectable tensor background, thermal inflation erases it at short scales \cite{Mendes:1998gr}. If this was the end of the story, one might regard it as an unfortunate side-effect of solving the moduli problem, but there is a final piece to the puzzle: thermal inflation ends via bubble nucleation, and the collisions between these bubbles as they percolate produces a significant stochastic gravitational wave background. We will see that this process may generate $\Omega_\mathrm{GW} h^2 (f \approx 1 \Hz) \sim 10^{-17}$, and is potentially detectable by BBO.

Thermal inflation is a complicated scenario, at least when compared to simple toy models of inflation. However, it relies on generic ingredients found in many supersymmetric or stringy models of the early universe; it is the intrinsic richness of these models that generates their complicated phenomenology. In particular, the moduli problem is robust enough to have survived 25 years. Conversely, thermal inflation does not require any exotic physics, unnatural parameter values or initial conditions, and the complexity of this scenario may in fact be typical of realistic models of the early universe. Finally, while the primordial gravitational background is erased, it would need to have been near the top of its plausible range to have been observable even without dilution, and furthermore in supersymmetric and stringy models the bottom of this range is arguably more natural \cite{Randall:1995dj,Kadota:2003fs,Kadota:2003tn}. The bubble collision signature is generic to thermal inflation and effectively expands the list of the models probed by BBO.

In Section 2 we summarize thermal inflation, and describe the process of bubble nucleation that occurs as it ends. In Section 3 we describe the production of gravitational waves  by bubble collisions and estimate the expected spectrum. In Section 4 we compute the suppression to the primordial inflationary background by a secondary period of inflation, combine this with the likely spectrum from bubble collisions, and assess the  prospects for detecting this signal in future observatories, the likelihood it can be distinguished from other stochastic astrophysical backgrounds, and the likely tensor signal  from the primary inflationary phase, which would be undiluted at CMB scales, or via pulsar timing experiments \cite{Allen:1996vm}. Finally we sum up in Section 5.

\section{Thermal inflation}

\subsection{Motivation}

The main motivation for thermal inflation \cite{Lyth:1995hj,Lyth:1995ka,Yamamoto:1985mb,Yamamoto:1985rd,Enqvist:1985kz,
Bertolami:1987xb,Ellis:1986nn,Ellis:1989ii,Randall:1994fr} is to solve the moduli problem \cite{Coughlan:1983ci,Banks:1993en,de Carlos:1993jw}, though it also solves the gravitino problem \cite{Khlopov:1984pf,Ellis:1984eq} and provides a mechanism for baryogenesis  \cite{Jeong:2004hy,Felder:2007iz,Stewart:1996ai,Kawasaki:2006py,Lazarides:1985ja,
Yamamoto:1986jw,Mohapatra:1986dg,Lazarides:1987yq}.
Irrespective of these uses, thermal inflation is sufficiently natural that it might occur anyway.

Moduli are scalar fields with Planckian vacuum expectation values, and hence gravitational strength interactions. Their potential arises due to supersymmetry breaking, and assuming supersymmetry breaking is transmitted to the observable sector via gravitational strength interactions, moduli have vacuum masses of order the soft supersymmetry breaking scale in the observable sector
\begin{equation} \label{ms}
m_\mathrm{moduli} \sim m_\mathrm{s} \sim 10^2 \textrm{ to } 10^3 \GeV.
\end{equation}
However, in the early universe, the finite energy density breaks supersymmetry. When $H \gtrsim m_\mathrm{s}$ this supersymmetry breaking dominates over the vacuum supersymmetry breaking and hence determines the moduli potential. When $H$ drops below $m_\mathrm{s}$ the moduli potential reduces to its vacuum form, but with the moduli typically displaced by a Planckian distance. The moduli then start oscillating with Planckian amplitude and immediately dominate the energy density of the universe. Thanks to the relatively low moduli mass and their very weak interactions, these oscillations persist beyond nucleosynthesis with disasterous consequences \cite{Coughlan:1983ci,Banks:1993en,de Carlos:1993jw}.

Inflation is typically invoked to rid the universe of unwanted relics, but one would expect an inflaton to have a mass $\gtrsim m_\mathrm{s}$ and so primordial inflation to occur at scales  $H \gtrsim m_\mathrm{s}$. The moduli are generated at $H \sim m_\mathrm{s}$, and to a lesser extent by any phase transition at $H \lesssim m_\mathrm{s}$. Thus one wants inflation at $H \ll m_\mathrm{s}$ to dilute the moduli, but it is very difficult to realize primordial inflation at these scales. On the other hand, thermal inflation \cite{Lyth:1995hj,Lyth:1995ka}  automatically occurs at $H \ll m_\mathrm{s}$. For a thermal inflation scale
\begin{equation}
V^{1/4} \sim 10^6 \textrm{ to } 10^7 \GeV
\end{equation}
thermal inflation provides enough dilution to rid the universe of moduli, but has a low enough scale not to regenerate them afterwards. Furthermore, as thermal inflation lasts about 10 $e$-folds or less, it does not destroy the primordial perturbations needed for structure formation.
While thermal inflation  provides a very natural solution to the moduli problem, it is incompatible with most baryogenesis scenarios since it will dilute baryons produced before it begins, and the reheat temperature after thermal inflation is very low ($\mathcal{O}(10\GeV)$).  Fortunately, a baryon asymmetry is almost automatically generated at the end of thermal inflation \cite{Jeong:2004hy,Felder:2007iz}, solving this difficulty.

\subsection{Particle physics model}
\label{section:particle}

\begin{figure}
\centering
\begin{tikzpicture}[x=0.1\textwidth,y=0.1\textwidth]
\draw[->] (0,0) -- (8,0) -- ++ (10pt,0pt) node[right]{$\phi$};
\draw[->] (0,0) -- (0,1.2) -- ++ (0pt,10pt) node[above]{$V$};
\draw (0,0) -- ++(0pt,-2pt);
\draw (0,0) -- ++(-2pt,0pt);
\path (0,0) -- ++(-2pt,-2pt) node[below left] {$0$};
\draw (6,0) -- ++(0pt,-2pt) node[below] {$\phi_0$};
\draw (0,1) -- ++(-2pt,0pt) node[left] {$V_\mathrm{TI}$};
\fill[blue] (0,0.8) circle(0.05);
\draw[red] plot[smooth] coordinates { (0,0.75) (0.025,0.758) (0.05,0.778) (0.075,0.807) (0.1,0.838) (0.125,0.867) (0.15,0.893) (0.175,0.914) (0.2,0.932) (0.25,0.957) (0.3,0.971) (0.4,0.984) (0.5,0.986) (0.6,0.984) };
\draw plot[smooth] coordinates { (0,1) (0.5,0.990) (1,0.958) (1.5,0.906) (2,0.834) (2.5,0.742) (3,0.633) (3.5,0.509) (4,0.377) (4.5,0.245) (5,0.126) (5.5,0.036) (5.75,0.010) (6,0) (6.25,0.011) (6.5,0.048) (6.75,0.115) (7,0.219) (7.25,0.366) (7.5,0.564) (7.75,0.819) (8,1.143) };
\end{tikzpicture}
\caption{ \label{fig:potential}
Thermal inflation occurs when a flaton $\phi$ is held at the origin by its finite temperature potential and $V_\mathrm{TI}$ dominates the energy density.
}
\end{figure}
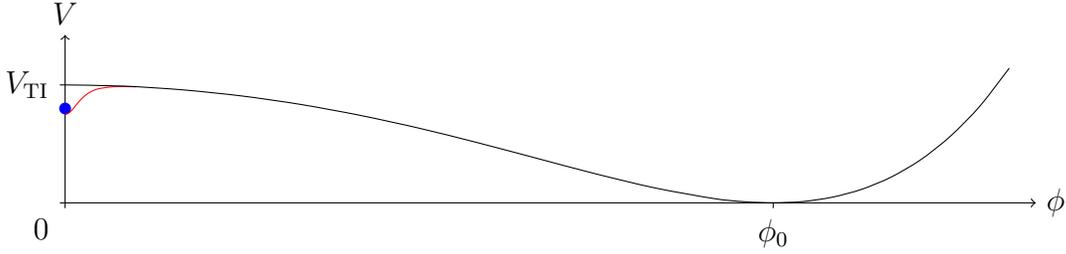

Like the Standard Model Higgs field, the flaton has a negative mass-squared at the origin
\begin{equation}
V(\phi) = V_\mathrm{TI} - \frac{1}{2} m_\phi^2 \phi^2 + \ldots
\end{equation}
with $m_\phi \sim m_\mathrm{s}$, the soft supersymmetry breaking mass scale.
Unlike the Standard Model Higgs field, but like many scalar field directions in the Minimal Supersymmetric Standard Model (MSSM) \footnote{For example, if the MSSM constraints $m^2_{H_u} + m^2_{H_d} + 2 |\mu|^2 > 2 \left| B \mu \right|$ and $( m^2_{H_d} + |\mu|^2 -  m^2_L ) ( m^2_L + m^2_{H_u} + |\mu|^2 ) > | B \mu |^2$ were not satisfied, then the MSSM would have a flaton.}, the flaton does not have a stabilizing $\phi^4$ term. Instead,  non-renormalizable higher order terms stabilize its potential at a large field value $\phi_0 \gg m_\phi$, as illustrated in Figure~\ref{fig:potential}.
The potential energy at the origin
\begin{equation}
V_\mathrm{TI} \sim m_\phi^2 \phi_0^2
\end{equation}
is tuned to give zero vacuum energy density.
For thermal inflation to solve the moduli problem, we require \cite{Lyth:1995ka}
\begin{equation}
V_\mathrm{TI}^{1/4} \sim 10^6 \textrm{ to } 10^7 \GeV \, ,
\end{equation}
corresponding to
\begin{equation}
\phi_0 \sim 10^{10} \textrm{ to } 10^{12} \GeV \, .
\end{equation}

The flaton is a Standard Model gauge singlet, since it would otherwise break the Standard Model gauge group at the scale $\phi_0$, but has unsuppressed Yukawa couplings to non-MSSM fields $\Psi$ and $\bar\Psi$, which acquire masses $M_\Psi \sim \phi_0$ once the flaton reaches the minimum of its potential.
$\Psi$ and $\bar\Psi$ are constrained by gauge coupling unification, and at least some of their Yukawa couplings to the flaton should be strong enough to drive the flaton's mass-squared negative at low energies, in the same way as the top Yukawa coupling drives the Higgs mass-squared negative at low energies.
Such a strong Yukawa coupling will be attracted to its renormalization group infrared fixed point, allowing one to relate its value at low energies to the strong gauge coupling and the choice of $\Psi$ and $\bar\Psi$ representations.
See \ref{appendix:couplings} for more details and numerical estimates for this Yukawa coupling.

The flaton's Yukawa terms couple it strongly to the thermal bath, so its finite temperature effective potential is
\begin{equation} \label{eqn:singlethermalpotential}
V(\phi,T) = V_\mathrm{TI} +
\left\{
\begin{array}{lcc}
- A T^4 + \frac{1}{2} \left( B T^2 - m_\phi^2 \right) \phi^2 + \ldots
& \textrm{for} & \phi \ll T \\
- \frac{1}{2} m_\phi^2 \phi^2 + \ldots
& \textrm{for} & \phi \gg T
\end{array}
\right.
\end{equation}
where the values of $A$ and $B$ depend on the details of the thermal bath and the flaton's coupling to the thermal bath, as we will see in \ref{appendix:potential}.
The potential in \eq{eqn:singlethermalpotential} has a global minimum at $\phi = \phi_0$ for $T \ll \phi_0$ and a local minimum at the origin for
\begin{equation}
T > T_{\mathrm{c}2} = \frac{m_\phi}{\sqrt{B}\,}
\end{equation}
where $T_{\mathrm{c}2}$ is the critical temperature for the would be second order phase transition.
However, we will find that thermal inflation always ends slightly \emph{before} the temperature drops far enough for the quadratic term to change sign. The field is thus trapped at the origin and the phase transition is first order, leading to bubble formation at a temperature $T_{\mathrm{c}1} > T_{\mathrm{c}2}$. Since $T_{\mathrm{c}1} \sim T_{\mathrm{c}2}$, when the difference between the temperatures is unimportant we will simply denote them by $T_\mathrm{c}$.

\subsection{Cosmological history}

\begin{figure}[tb]
\centering
\tikzstyle{moduli}=[black,very thick]
\tikzstyle{potential}=[green,very thick]
\tikzstyle{flaton}=[blue,very thick]
\tikzstyle{radiation}=[red,very thick]
\begin{tikzpicture}[x=0.015\textwidth,y=0.01\textwidth]
\fill[black!10] (0,22) rectangle (52,10);
\fill[green!10] (0,10) rectangle (52,0);
\fill[blue!10] (0,0) rectangle (52,-12);
\fill[red!10] (0,-12) rectangle (52,-27);
\draw[->] (52,34) -- (0,34) -- ++ (-10pt,0pt) node[left]{$\Omega$};
\draw[->] (0,34) -- (0,-27) -- ++ (0pt,-10pt) node[below]{$t$};
\draw[->] (52,34) -- (52,-27) -- ++ (0pt,-10pt) node[below]{$\ln a$};
\draw[dotted] (0,22) -- (52,22);
\draw[dotted] (0,10) -- (52,10);
\draw[dotted] (0,0) -- (52,0);
\draw[dotted] (0,-12) -- (52,-12);
\draw[dotted] (12,34) -- (12,-27);
\draw[dotted] (36,34) -- (36,-27);
\draw[dotted] (30,34) -- (30,-27);
\draw (0,34) -- ++(0pt,2pt) node[above]{1};
\draw (12,34) -- ++(0pt,2pt) node[above]{$e^{-12}$};
\draw (36,34) -- ++(0pt,2pt) node[above]{$e^{-36}$};
\draw (30,34) -- ++(0pt,2pt) node[above]{$e^{-30}$};
\draw (52,34) -- ++(0pt,2pt) node[above]{$e^{-52}$};
\draw (0,22) -- ++(-2pt,0pt) node[left]{$t_\mathrm{m}$};
\draw (0,10) -- ++(-2pt,0pt) node[left]{$t_\mathrm{b}$};
\draw (0,0) -- ++(-2pt,0pt) node[left]{$t_\mathrm{c}$};
\draw (0,-12) -- ++(-2pt,0pt) node[left]{$t_\mathrm{d}$};
\draw (52,22) -- ++(2pt,0pt) node[right]{-22};
\draw (52,10) -- ++(2pt,0pt) node[right]{-10};
\draw (52,0) -- ++(2pt,0pt) node[right]{0};
\draw (52,-12) -- ++(2pt,0pt) node[right]{12};
\path (0,30) -- node[gray]{? GUT INFLATION + RADIATION DOMINATION ?} (52,30);
\path (0,26) -- node[gray]{? MODULAR INFLATION ?} (52,26);
\path (0,22) -- node[moduli]{$H_\mathrm{m} \sim 10^2 \textrm{ to } 10^3 \GeV$} (30,22);
\path (0,16) -- node[moduli]{MODULI DOMINATION} (30,16);
\path (0,5) -- node[potential]{THERMAL INFLATION} (30,5);
\path (30,5) -- node[potential]{$V_\mathrm{TI}^{1/4} \sim 10^6 \textrm{ to } 10^7 \GeV$} (52,5);
\path (30,0) -- node[radiation]{$T_\mathrm{c} \sim 10^2 \textrm{ to } 10^3 \GeV$} (52,0);
\path (0,-1) -- node[flaton]{baryogenesis} (30,-1);
\path (0,-6) -- node[flaton]{FLATON DOMINATION} (30,-6);
\path (0,-12) -- node[radiation]{$T_\mathrm{d} \sim 1 \textrm{ to } 10^2 \GeV$} (30,-12);
\path (0,-15.75) -- node[radiation]{dark matter generated} (30,-15.75);
\path (0,-19.5) -- node[radiation]{RADIATION DOMINATION} (30,-19.5);
\path (0,-23.25) -- node[radiation]{nucleosynthesis} (30,-23.25);
\draw[radiation] (0,22) -- (12,10) -- (52,0);
\draw[moduli] (0,22) -- (0,10) -- (30,0) -- (30,-12) -- (15,-27);
\draw[potential] (36,22) -- (0,10) -- (0,0);
\draw[flaton] (0,0) -- (0,-12);
\draw[radiation] (0,-12) -- (0,-27);
\end{tikzpicture}
\caption{ \label{history}
History of the universe with thermal inflation.
The fractional density $\Omega_X \equiv \rho_X/\rho$ of
moduli \tikz{\path (0em,0ex) rectangle (1em,1ex); \draw[moduli] (0em,0.5ex) -- (1em,0.5ex);},
potential \tikz{\path (0em,0ex) rectangle (1em,1ex); \draw[potential] (0em,0.5ex) -- (1em,0.5ex);},
flaton \tikz{\path (0em,0ex) rectangle (1em,1ex); \draw[flaton] (0em,0.5ex) -- (1em,0.5ex);} and
radiation \tikz{\path (0em,0ex) rectangle (1em,1ex); \draw[radiation] (0em,0.5ex) -- (1em,0.5ex);}
is plotted against the number of $e$-folds of expansion $\ln a$.
}
\end{figure}
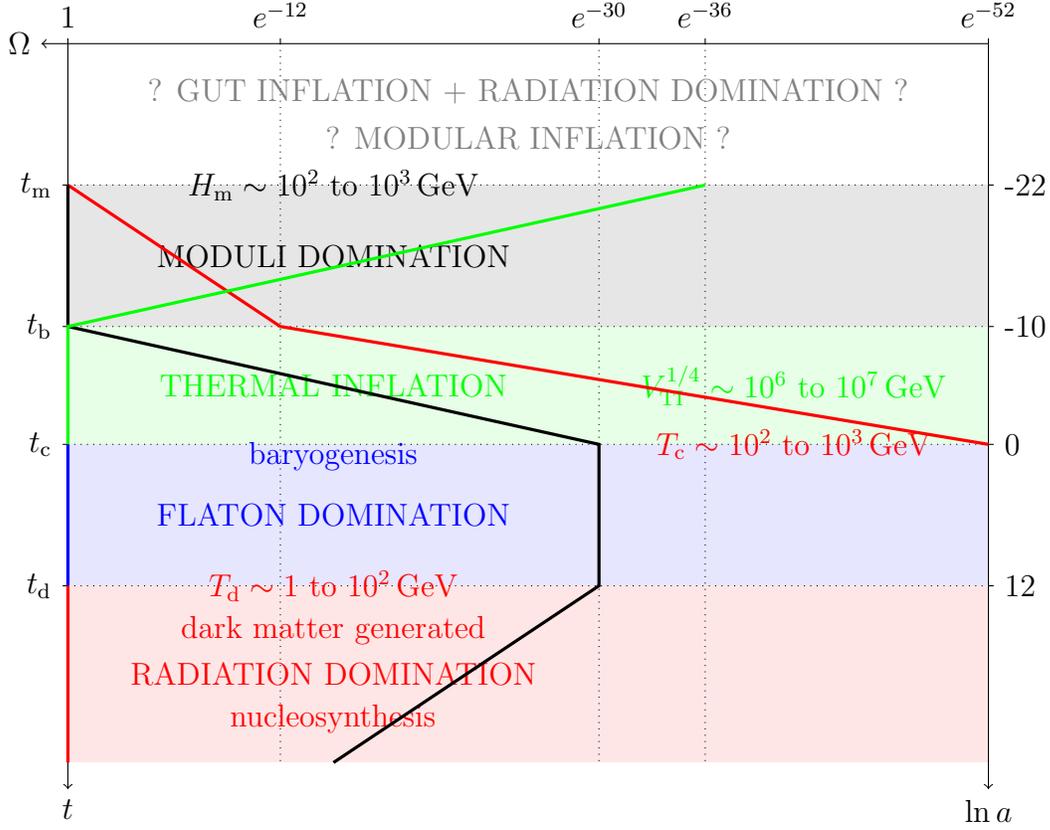

The overall history of a universe which undegoes thermal inflation is summarized in Figure~\ref{history}. The primordial inflationary phase, which lays down the perturbation spectra, ends at some time $t \leq t_\mathrm{m}$. In supersymmetric theories, the moduli are produced at the time $t = t_\mathrm{m}$ when $H \sim m_\mathrm{s}$. At this time the moduli potential switches from its finite energy density form to its vacuum form, leaving the moduli oscillating with Planckian amplitude and $\rho_\mathrm{moduli} \sim \rho_\mathrm{rad}$.  However, the moduli quickly dominate, as $\rho_\mathrm{moduli} \propto a^{-3}$ and $\rho_\mathrm{rad} \propto a^{-4}$.

Thermal inflation begins at $t_\mathrm{b}$,  at which point  $\rho_\mathrm{moduli} \sim V_\mathrm{TI}$ and
\begin{equation}
\frac{\rho_\mathrm{moduli}}{\rho_\mathrm{rad}}
\sim \frac{a_\mathrm{b}}{a_\mathrm{m}}
\sim \left( \frac{H_\mathrm{m}}{H_\mathrm{b}} \right)^{2/3}
\sim \left( \frac{m_\mathrm{s}^2 M_\mathrm{Pl}^2}{V_\mathrm{TI}} \right)^{1/3}
\end{equation}
where $M_\mathrm{Pl} \equiv 1/\sqrt{8\pi G} = 2.4 \times 10^{18} \GeV$, and
\begin{equation} \label{Ti}
T_\mathrm{b}
\sim \left( \frac{\rho_\mathrm{rad}}{\rho_\mathrm{moduli}} \right)^{1/4} V_\mathrm{TI}^{1/4}
\sim \left( \frac{V_\mathrm{TI}^2}{m_\mathrm{s} M_\mathrm{Pl}} \right)^{1/6} \, .
\end{equation}
Thermal inflation ends at $t_\mathrm{c}$ when
\begin{equation} \label{Tf}
T = T_\mathrm{c} \sim m_\phi \, .
\end{equation}
During thermal inflation, the temperature of the thermal bath drops as $T \propto a^{-1}$, assuming the effective number of degrees of freedom does not change significantly.  Therefore, from \eqs{Ti}{Tf}, thermal inflation lasts for less than 10 $e$-folds \footnote{This estimate could be increased by about 5 $e$-folds in the baryogenesis scenario of Refs.~\cite{Jeong:2004hy,Felder:2007iz}.}, since
\begin{equation}
N_\mathrm{TI}
\simeq \ln \left( \frac{T_\mathrm{b}}{T_\mathrm{c}} \right)
\simeq \frac{1}{6} \ln \left( \frac{\phi_0^4}{m_\phi^2 m_\mathrm{s} M_\mathrm{Pl}} \right)
\simeq 6 \textrm{ to } 9 \, .
\end{equation}
After thermal inflation, the universe is dominated by the flaton and $\rho_\phi \propto a^{-3}$ \footnote{Preheating may alter this equation of state slightly.}. Finally, the flaton decays at $t = t_\mathrm{d}$ leaving a radiation dominated universe at temperature
\begin{equation}
T_\mathrm{d} \sim 1 \textrm{ to } 10^2 \GeV  \, .
\end{equation}
At this point, the stage is set for the standard hot Big Bang.

\subsection{The first order phase transition at the end of thermal inflation}
\label{section:transition}

The origin is always only a local minimum of the flaton's finite temperature effective potential, all the way down to the temperature $T_\mathrm{c2}$, so the phase transition that ends thermal inflation will be first order. We therefore apply the thermal tunnelling formalism of Refs.~\cite{Linde:1980tt,Linde:1981zj,Linde:2005ht}. After nucleation, there is a very large pressure difference across the bubble wall due to the deep vacuum of the flaton field, hence we expect the bubble wall propagation to proceed as a detonation \cite{Steinhardt:1981ct}, and the walls rapidly accelerate to relativistic speeds. Finally, the bubbles will percolate, ending thermal inflation.

The bubble nucleation rate is
\begin{equation} \label{decayrate}
\Gamma(T) \sim T^4 \exp{ \left[ - \frac{S_3(T)}{T} \right] } \, ,
\end{equation}
where $S_3(T)$ is the energy of a bubble.
The temperature $T_{\mathrm{c}1}$ at which the dominant bubbles nucleate is given by
\begin{equation} \label{nucleation}
\fn{\Gamma}{T_{\mathrm{c}1}} \sim \left[ \frac{ \fn{\dot\Gamma}{T_{\mathrm{c}1}} }{ \fn{\Gamma}{T_{\mathrm{c}1}} } \right]^4 \, ,
\end{equation}
see \ref{appendix:tunnelling}. The timescale of the transition, from nucleation to percolation, and hence the characteristic size of the bubbles at percolation, is given by the important parameter $\beta^{-1}$
\begin{equation} \label{eqn:timescale}
\frac{\beta}{H_\mathrm{c}}
\equiv \left. \frac{\dot{\Gamma}}{H \Gamma} \right|_{T_{\mathrm{c}1}}
\simeq - \left[ \frac{1}{H} \frac{d}{dt} \left( \frac{S_3}{T} \right) \right]_{T_{\mathrm{c}1}}
\simeq \left[ T \frac{d}{dT} \left( \frac{S_3}{T} \right) \right]_{T_{\mathrm{c}1}}  \, ,
\end{equation}
where
\begin{equation} \label{hubble}
H_\mathrm{c} = \sqrt{ \frac{V_\mathrm{TI}}{3 M_\mathrm{Pl}^2} }
\end{equation}
is the Hubble parameter at the time of nucleation and percolation. The exact value of $\beta$ depends on the strength of the phase transition, which in turn depends crucially on the details of the thermal bath which holds the flaton at the top of its potential. In \ref{appendix:estimation}, we use the constraints on the flaton's Yukawa couplings, obtained in \ref{appendix:couplings}, to derive estimates for $\beta/H_\mathrm{c}$, finding
\begin{equation} \label{eqn:betaest}
\frac{\beta}{H_\mathrm{c}} \sim 10^3 \textrm{ to } 10^4 \, .
\end{equation}

\section{Gravitational wave production during bubble percolation}
\label{sect:GWprod}

Gravitational waves are generated when the bubble walls collide  \cite{Kosowsky:1992rz}, and there may be an additional contribution from turbulence \cite{Kamionkowski:1993fg}. %
The ratio
\begin{equation}
\frac{\beta}{H_\mathrm{c}} \equiv \left. \frac{\dot\Gamma}{H \Gamma} \right|_{T_{\mathrm{c}1}}
\end{equation}
determines both the peak frequency and amplitude of the gravitational wave spectrum generated by bubble percolation. Our result for  $\beta/H_\mathrm{c}$, \eq{eqn:betaest}, tells us that the bubble radii at percolation are much smaller than the Hubble radius. This justifies the use of the gravitational wave spectrum derived numerically in Ref.~\cite{Kamionkowski:1993fg} which assumed that the gravitational waves are generated in an effectively flat background. Physically,  $\beta/H_\mathrm{c}$ determines the peak frequency because the size of the bubbles determines the characteristic lengthscale of the quadrupole moment that generates the gravitational waves, and hence the frequency at which we observe them today.

The peak amplitude is a little more subtle. The total energy radiated in gravity waves, per bubble, per frequency interval $d\omega$ and solid angle $\tilde{\Omega}$ is \cite{WeinbergGRBook}
\begin{equation}
\frac{dE_{gw}}{d\omega d\tilde{\Omega}} = 2G\omega^2 \Lambda_{ij,lm}({\bf{k}})T_{ij}^*({\bf{k}},\omega)T_{lm}(\bf{k},\omega)
\end{equation}
where $\Lambda_{ij,lm}$ is the projection tensor that picks up the traceless and transverse component of the source stress tensor, with $\hat{\bf{k}}\equiv {\bf{k}}/|k|$
\begin{equation}
\Lambda_{ij,lm} = \delta_{ij}\delta_{jm}-2\bvect{j}\bvect{m}\delta_{il}+\frac{1}{2}\bvect{i}\bvect{j}\bvect{l}\bvect{m}-\frac{1}{2}\delta_{ij}\delta_{lm}+\frac{1}{2}\delta_{ij}\bvect{l}\bvect{m}+\frac{1}{2}\delta_{lm}\bvect{i}\bvect{j}.
\end{equation}
$T_{ij}({\bf{k}},\omega)$ is the fourier transform of the source stress tensor for each nucleating spherical bubble with size $R$ \cite{Kamionkowski:1993fg}
\begin{equation}
T_{ij}({\bf{k}},\omega) =  \frac{1}{6\pi}\int_0^{\infty} dt~ e^{i\omega t} \int d\tilde{\Omega}~ R^3 \rho_v \kappa \bvectx{i}\bvectx{j}.
\end{equation}
$\kappa \leq 1$ is the efficiency factor for the conversion of the vacuum energy $\rho_v$ into the kinetic energy of the bubbles.
This imply that, per bubble, $dE_{gw}/d\omega \propto (R^3 \kappa \rho_v)^2$. Notice that it is a product of the vacuum energy density which can be traced back to the fact that the quadrupole moment of a gravitational source is proportional to the source energy density. Meanwhile, the total energy is simply $E_v \propto R^3 \rho_v$, and substituting the lengthscale of the relativistic bubble $R\propto \beta^{-1}$ and integrating over the peak production frequency $\omega \propto \beta$ we get
\begin{equation}
\frac{E_\mathrm{GW}}{E_\mathrm{total}} = \Omega_\mathrm{GW} \propto \kappa^2 \left(\frac{H}{\beta}\right)^2.
\end{equation}
recalling that $H^2 \propto \rho_v$.
This relation was first derived semi-analytically and checked numerically in references \cite{Kosowsky:1992vn,Kosowsky:1992rz}. In particular,  the \emph{total} energy in gravitational waves generated by thin-wall bubble collisions is well-fitted by the heuristic formula \cite{Kamionkowski:1993fg}
\begin{equation}
\frac{E_\mathrm{GW}}{E_\mathrm{total}} \approx 0.07\kappa^2 \left(\frac{H}{\beta}\right)^2\left(\frac{\alpha}{1+\alpha}\right)^2\left(\frac{v^3}{0.24+v^3}\right)
\end{equation}
where $\alpha$ is the ratio of vacuum energy to thermal energy before the phase transition occurs
\begin{equation}
\alpha = \frac{V_\mathrm{TI}}{\rho_\mathrm{rad}(T_\mathrm{c})} \sim 10^{15} \, ,
\end{equation}
$v$ is the wall velocity, and as discussed in Section~\ref{section:transition} we expect $v \simeq 1$, and since the thermal energy is negligible we expect $\kappa \simeq 1$.

In order to assess the observability of this signal by a given detector   we  need the spectrum
\begin{equation}
\Omega_\mathrm{GW}(\omega)\equiv  \frac{d E_\mathrm{GW}/E_\mathrm{total}}{d\ln\omega}.
\end{equation}
This spectrum has been numerically computed for transitions of the sort we are concerned with here \cite{Kosowsky:1992vn,Kamionkowski:1993fg}. In particular, looking at Figure~7 of Ref.~\cite{Kamionkowski:1993fg} with $v=1$, the peak value for $\Omega_{\mathrm{GW}}$ is
\begin{equation} \label{ogwp}
\fn{\Omega_\mathrm{GW}}{\omega \approx \beta} = 0.03 \times \frac{H^2}{\beta^2}\kappa^2 \frac{\alpha^2}{(1+\alpha)^2}.
\end{equation}
In other words, the peak power is at a characteristic frequency $\omega = \beta$. This is then rescaled by the subsequent cosmological expansion,  so the frequency today is
\begin{eqnarray}f(t_0)& = & \frac{H_\mathrm{c}}{2\pi} \left( \frac{\beta}{H_\mathrm{c}} \right) \left( \frac{a_\mathrm{c}}{a_0} \right)
\\ \label{peakfrequency}
& \simeq & 0.7 \Hz
\left( \frac{\beta/H_\mathrm{c}}{1000} \right)
\left( \frac{V_\mathrm{TI}^{1/4}}{10^6 \,\mathrm{GeV}} \right)^{2/3}
\left( \frac{T_\mathrm{d}}{10^2 \,\mathrm{GeV}} \right)^{1/3}
\left( \frac{V_\mathrm{TI}^{1/3} a_\mathrm{c}}{\rho_\mathrm{d}^{1/3} a_\mathrm{d}} \right).
\label{eqn:peakfreq}
\end{eqnarray}
The last factor is unity if the universe is dominated by non-relativistic flaton matter after thermal inflation, but bubble collisions and other preheating effects would be expected to heat up the flaton matter at least initially, making this factor greater than unity. However, these effects are very difficult to estimate so we assume the most conservative value of unity. Assuming a more favorable set of thermal inflation  parameters for the remaining factors ($\beta/H_\mathrm{c} = 10^3$, $V_\mathrm{TI}^{1/4} = 10^6 \GeV$, $T_d = 10^2 \GeV$), the peak frequency today is about $0.7 \Hz$, well within the spectral range of future space-based gravitational wave detectors such as LISA \cite{LISA} and BBO \cite{BBO}.
Similarly, we can map the power spectrum into its present day form
\begin{eqnarray}
\lefteqn{
\fn{\Omega_\mathrm{GW}}{f_\mathrm{peak},t_0} h^2
= \fn{\Omega_\mathrm{GW}}{f_\mathrm{c},t_\mathrm{c}} h^2 \left( \frac{a_\mathrm{c}}{a_0} \right)^4 \left( \frac{H_\mathrm{c}}{H_0} \right)^2
} \\ \label{eqn:energydensity}
& \simeq & 5 \times 10^{-18}
\left( \frac{\beta/H_\mathrm{c}}{1000} \right)^{-2}
\left( \frac{V_\mathrm{TI}^{1/4}}{10^6 \,\mathrm{GeV}} \right)^{-4/3}
\left( \frac{T_\mathrm{d}}{10^2 \,\mathrm{GeV}} \right)^{4/3}
\left( \frac{V_\mathrm{TI}^{1/3} a_\mathrm{c}}{\rho_\mathrm{d}^{1/3} a_\mathrm{d}} \right)^4 \, .
\end{eqnarray}
With the same parameters as above, we find a peak gravitational wave power of  $\fn{\Omega_\mathrm{GW}}{f} h^2 \sim 5 \times 10^{-18}$, which is close to BBO's design sensitivity, and well below that of LISA.

This amplitude is lower than the $\fn{\Omega_\mathrm{GW}}{f} h^2 \sim 10^{-11}$ seen after preheating  or some turbulent processes \cite{GarciaBellido:2007af,Easther:2006vd,Easther:2006gt,Khlebnikov:1997di,Dufaux:2007pt,
GarciaBellido:2007dg,Grojean:2006bp,Randall:2006py}.  The bubbles are initially small, since $\beta/H_\mathrm{c} \sim 1000$,  which leads to a slight suppression of power, via \eq{eqn:energydensity}. More importantly  the period of flaton matter domination following bubble nucleation dilutes the power by a redshift factor of $a_\mathrm{c}/a_\mathrm{d} \sim 10^{-5}$, relative to what would be seen if the universe became radiation dominated immediately after the bubble collisions. On the other hand, we have not considered gravitational waves sourced by turbulent mixing well after the bubble collisions, which is a potential source of additional power -- especially as these gravitational waves would be less diluted by the flaton dominated epoch. \eq{eqn:energydensity} is thus a \emph{lower bound} for the gravitational wave power. Further, much of the energy of the universe is contained in the walls, which are moving at relativistic speeds. This should help to make the initial preheating of the flaton very efficient, and the details of this will need to be carefully studied in order to determine the gravitational wave spectrum generated in the post-collision universe.

In addition, Caprini, Durrer and Servant \cite{Caprini:2007xq} have recently given an elegant  analytic discussion of gravitational wave production during bubble collisions. The amplitude and shape they obtain for the resulting spectrum is broadly similar to the numerical results of \cite{Kamionkowski:1993fg} (note that \cite{Caprini:2007xq} assumes there is no matter-dominated phase after percolation), adding support to our estimate.  They also note that bubble collisions can in fact be a sub-dominant source of gravitational waves following a phase transition, and a full analysis of this signal will require further study.

\section{Forecasts for Observations} \label{sect:forecasts}

If a CMB B-mode is sourced by a primordial spectrum of gravitational waves, the properties of this signal do not depend strongly on the details post-inflationary physics, since the relevant modes are outside the horizon until after matter-radiation equality.   Conversely, direct detection experiments are sensitive to modes  with much smaller wavelengths, and their current amplitude can depend strongly on the expansion history of the post-inflationary universe \cite{Boyle:2005se,Boyle:2007zx}. In particular, since gravitational waves scale like radiation with $\rho_\mathrm{GW} \propto 1/a^4$, any phase in which  the universe is dominated by a component whose density scales away less rapidly than radiation will suppress the gravitational background on sub-horizon scales, as first discussed by \cite{Mendes:1998gr}.  Looking at the cosmological history implied by thermal inflation, we find three distinct epochs during which the {\em short wavelength\/} primordial gravitational wave background can be diluted:  moduli matter domination,  thermal inflation itself, and flaton domination after thermal inflation.

\begin{figure}[t]
\centering
\tikzstyle{hubble}=[black,very thick]
\tikzstyle{mode}=[black,very thick,dotted]
\tikzstyle{gw}=[violet]
\tikzstyle{moduli}=[black]
\tikzstyle{potential}=[green]
\tikzstyle{flaton}=[blue]
\tikzstyle{radiation}=[red]
\begin{tikzpicture}[x=0.0125\textwidth,y=0.0125\textwidth]
\fill[black!10] (-33,22) rectangle (38,10);
\fill[green!10] (-33,10) rectangle (38,0);
\fill[blue!10] (-33,0) rectangle (38,-12);
\fill[red!10] (-33,-12) rectangle (38,-22);
\draw[->] (-33,32) -- (38,32) -- ++ (10pt,0pt) node[right]{$\ln\lambda$};
\draw[->] (-33,32) -- (-33,-22) -- ++ (0pt,-10pt) node[below]{$t$};
\draw[->] (38,32) -- (38,-22) -- ++ (0pt,-10pt) node[below]{$\ln a$};
\draw[dotted] (-33,22) -- (38,22);
\draw[dotted] (-33,10) -- (38,10);
\draw[dotted] (-33,0) -- (38,0);
\draw[dotted] (-33,-12) -- (38,-12);
\draw[dotted] (-7,32) -- (-7,-22);
\draw[dotted] (0,32) -- (0,-22);
\draw[dotted] (18,32) -- (18,-22);
\draw (-7,32) -- ++(0pt,2pt) node[above]{-7};
\draw (0,32) -- ++(0pt,2pt) node[above]{0};
\draw (18,32) -- ++(0pt,2pt) node[above]{18};
\draw (-33,22) -- ++(-2pt,0pt) node[left]{$t_\mathrm{m}$};
\draw (-33,10) -- ++(-2pt,0pt) node[left]{$t_\mathrm{b}$};
\draw (-33,0) -- ++(-2pt,0pt) node[left]{$t_\mathrm{c}$};
\draw (-33,-12) -- ++(-2pt,0pt) node[left]{$t_\mathrm{d}$};
\draw (38,22) -- ++(2pt,0pt) node[right]{-22};
\draw (38,10) -- ++(2pt,0pt) node[right]{-10};
\draw (38,0) -- ++(2pt,0pt) node[right]{0};
\draw (38,-12) -- ++(2pt,0pt) node[right]{12};
\path (-33,28.7) -- node[gray]{? GUT INFLATION + RADIATION DOMINATION ?} (38,28.7);
\path (-33,25.3) -- node[gray]{? MODULAR INFLATION ?} (38,25.3);
\path (0,22) -- node[moduli]{$H_\mathrm{m} \sim 10^2 \textrm{ to } 10^3 \GeV$} (38,22);
\path (0,16) -- node[moduli]{MODULI DOMINATION} (38,16);
\path (-33,5) -- node[potential]{THERMAL INFLATION} (-7,5);
\path (9,5) -- node[potential]{$V_\mathrm{TI}^{1/4} \sim 10^6 \textrm{ to } 10^7 \GeV$} (38,5);
\path (18,0) -- node[radiation]{$T_\mathrm{c} \sim 10^2 \textrm{ to } 10^3 \GeV$} (38,0);
\path (-33,-6) -- node[flaton]{FLATON DOMINATION} (-7,-6);
\path (-33,-12) -- node[radiation]{$T_\mathrm{d} \sim 1 \textrm{ to } 10^2 \GeV$} (-3.5,-12);
\path (-33,-17) -- node[radiation]{RADIATION DOMINATION} (0,-17);
\draw[hubble] (-33,22) -- node[above,sloped]{Hubble radius} (0,10) -- (0,0) -- (18,-12) -- (38,-22);
\draw[mode] (-9,22) -- (35,-22) node[below]{1};
\draw[mode] (-14,22) -- node[above,sloped]{comoving modes} (30,-22) node[below]{2};
\draw[mode] (-19,22) -- (25,-22) node[below]{3};
\draw[mode] (-24,22) -- (20,-22) node[below]{4};
\draw[mode] (-29,22) -- (15,-22) node[below]{5};
\fill[gw] (-8,0) -- (-6,0) -- (-5.5,-0.5) -- (-7.5,-0.5) node[left,gw]{peak gravitational} node[below left,gw]{wave generation};
\end{tikzpicture}
\caption{ \label{fig:scaleplot}
We illustrate the evolution of five representative gravitational wave modes (dotted lines) and the Hubble radius (solid line) in a universe with thermal inflation. Physical wavelength $\lambda$ is plotted against the scale factor $a$. The axes are natural logarithm scales normalized so that the scale factor and the Hubble radius are unity at the end of thermal inflation. For $\beta/H_\mathrm{c} \sim 10^3$, the bubbles are about $e^7$ times smaller than the horizon. The violet patch indicates the times and scales at which bubble collisions generate gravitational waves. The wavelength of Mode~1 is large enough to keep it outside the horizon until well after thermal inflation, and it is thus unaffected by the turmoil within the horizon. Mode~2 enters the horizon during moduli matter domination, and its amplitude is diluted during this phase and during thermal inflation, but it freezes out when it re-exits the horizon. Mode 3 resembles Mode 2 except that it reenters the horizon during flaton matter domination, and is thus suppressed a second time. Mode~4 remains within the horizon during thermal inflation, and is this maximally suppressed.   Conversely Mode~5 also remains within the horizon, but is short enough to be sourced by the bubble collisions.
}
\end{figure}
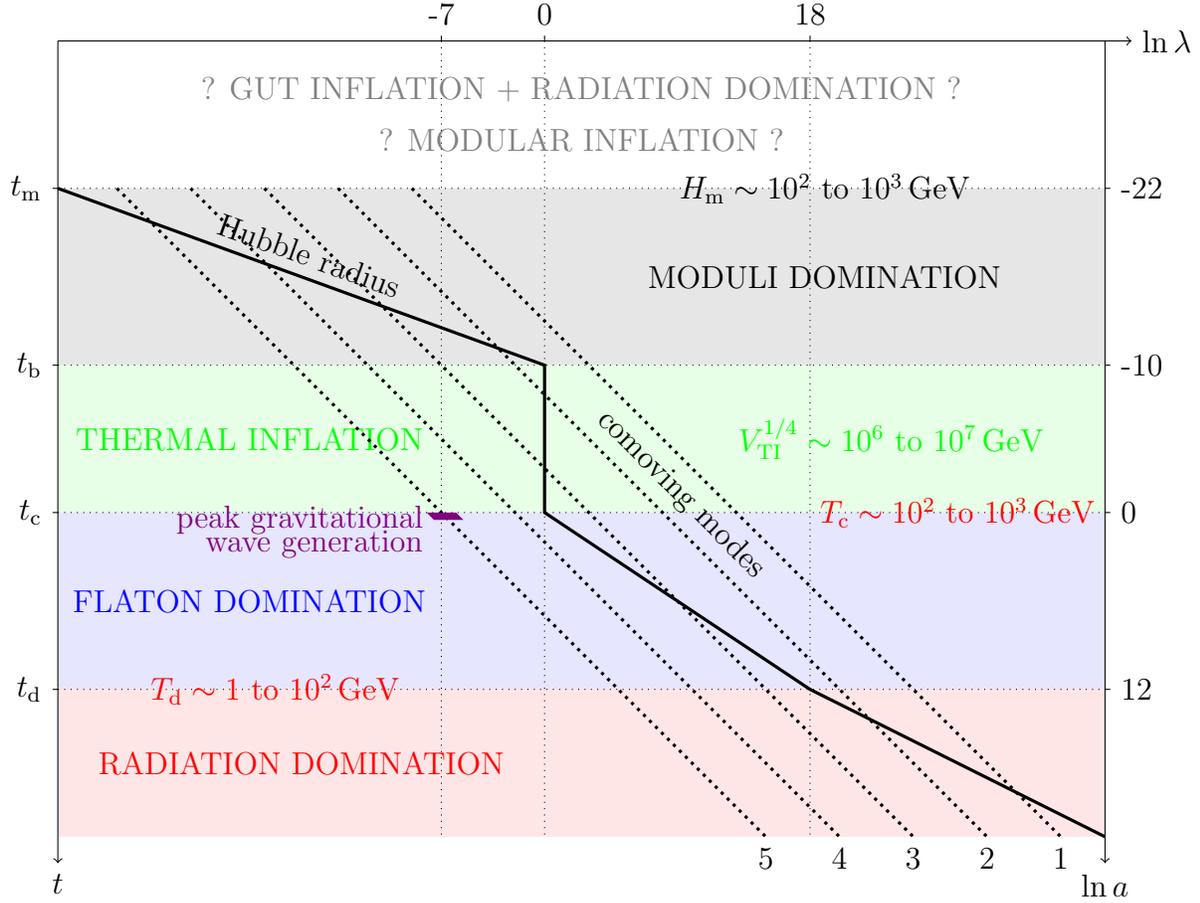

\begin{figure}[t]
\myfigure{6in}{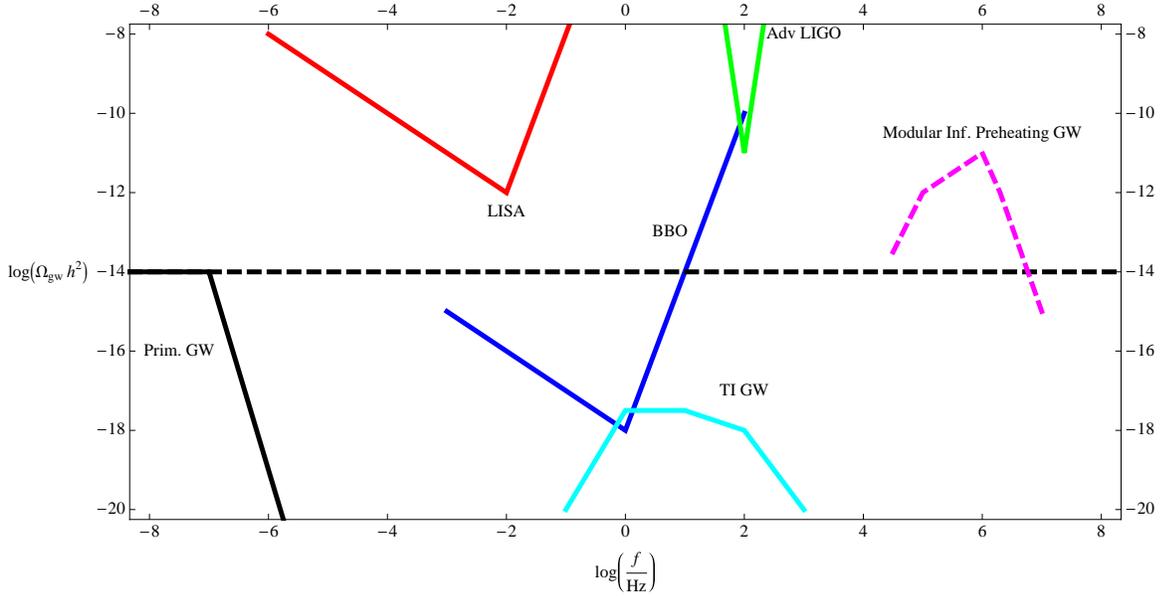}
\caption{ \label{fig:thermalplot}
The expected gravitational spectrum for a universe with a thermal inflationary phase, with projected sensitivities of future planned experiments. The black line shows the upper bound on the primordial gravitational wave spectrum from CMB data at the time of writing.  For reasonable parameter values, thermal inflation leads to a suppression of the primordial signal for $f >10^{-6} \Hz$, rendering it invisible to direct detection experiments.  The cyan line indicates the spectrum generated by bubble collisions at the end of thermal inflation, using the fiducial values of $\beta/H \approx 1000$, $V_\mathrm{TI}^{1/4} \approx 10^6 \GeV$ and $T_\mathrm{d} \approx 10^2 \GeV$. The magenta line is the expected GW spectrum from parameteric resonance \cite{Finelli:1998bu,Kofman:1997yn} after modular inflation \cite{Randall:1995dj,Kadota:2003fs,Kadota:2003tn}. It gets wiped out by thermal inflation.
}
\end{figure}

In Figure~\ref{fig:scaleplot} we show the history of five representative modes.   All modes reenter the horizon after the primary phase of inflation, with an amplitude determined by standard inflationary perturbation theory \cite{Mukhanov:1990me}, $\Omega_\mathrm{GW}^\mathrm{inf}(k)\propto H_\mathrm{inf}^2$. During moduli matter domination a mode inside the horizon will be diluted by a factor of   $a^{-1}$. As soon as thermal inflation kicks in the suppression factor is $a^{-4}$, dropping back to zero if the mode re-exits the horizon and re-freezes. Modes inside the horizon are further diluted by a factor of $a^{-1}$ during the flaton dominated phase, but this extra suppression is largely academic,  other than for modes which are sourced by the bubble collisions.   A schematic plot of the final spectrum is given in Figure~\ref{fig:thermalplot}. Note that the primordial gravitational wave spectrum is unsuppressed on long scales, so the constraint on the inflationary scale implied by the upper bound on the CMB B-mode is  still valid \cite{Spergel:2006hy}.

For a fixed value of $\Omega_\mathrm{GW}(f)$, gravitational waves are easier to
detect at long wavelengths, as the required strain-sensitivity scales as $f^3$ \cite{Maggiore:2000gv}.   Via \eq{eqn:peakfreq}, we see that the peak frequency scales as $H_\mathrm{c}^{1/3}$, and a lower thermal inflation scale leads to a redder peak, and thus a more detectable signal.  Further, for fixed $T_\mathrm{d}$, lowering the thermal inflation scale shortens the flaton dominated period, which, from \eq{eqn:energydensity}, increases the present day amplitude by a factor $H_\mathrm{c}^{-2/3}$. On the other hand, if $T_d$ is increased,  the amplitude is boosted by a factor of $T_d^{4/3}$, while the peak frequency grows as $T_d^{1/3}$. Thus, \emph{low scale} thermal inflation with a short period of flaton matter domination is the optimal prescription for generating an observable spectrum of gravitational waves during the  bubble nucleation phase. In terms  of the thermal inflation parameter space, this situation is equivalent to having a small value of  $\phi_0$.

For generic parameter values, this signal will peak at around frequencies of $1\Hz$, as shown in   Figure~\ref{fig:thermalplot}.   This corresponds to a wavelength on the order of $10^5\km$, which is characteristic of BBO or DECIGO style space-based proposals \cite{Seto:2001qf,Boyle:2005se}.  Moreover, the strong spectral dependence of the bubble collision signal will differentiate it from that the featureless primordial spectrum.

\section{Discussion}

This analysis draws on a  number of apparently disparate topics in theoretical cosmology. Firstly, it looks carefully at thermal inflation, considering both its expansion history and the properties of the bubbles created by the phase transition that marks its end. Historically, the primary motivation for considering thermal inflation is that it provides a natural solution to the moduli problem, which is endemic to supersymmetric models and arises from the production of a condensate of long-lived moduli particles which either disrupt nucleosynthesis as they decay or massively overclose the universe. Thermal inflation dilutes these unwanted particles to acceptable levels.   However, we show that thermal inflation also dilutes the {\em primordial\/} gravitational wave background at small scales, rendering it effectively unobservable by direct detection experiments such as BBO, but preserving any signal that is present in the CMB B-mode.  The detailed dependence of the BBO signal on the effective equation of state in the primordial universe has been discussed elsewhere (e.g.\  \cite{Boyle:2005se}) but the dilution factor implied by thermal inflation would effectively erase any primordial signal.

Conversely, thermal inflation ends via a first order phase transition, so the percolation of the resulting bubbles can generate a new background of gravitational waves.  This spectrum is confined to a few decades in frequency -- unlike the scale-free inflationary background -- and is potentially detectable by BBO style experiments.  While the detailed mechanism is different, this signal has much in common with the gravitational wave background that can be generated during preheating or parametric resonance \cite{GarciaBellido:2007af,Easther:2006gt,Easther:2006vd,Khlebnikov:1997di,
Dufaux:2007pt,GarciaBellido:2007dg,Easther:2007vj}.  Further study will be needed to see if the characteristic spectra can be easily distinguished from one another.

The analysis here only considers the background generated by the bubble collisions themselves -- a period of turbulent mixing after the bubble collisions would also source gravitational waves, and this signal has not yet been computed. It has the potential to significantly enhance the gravitational wave signal, since the bubble nucleation phase is followed by a period of flaton domination, mimicking a matter dominated universe, and diluting the background produced by colliding bubbles. If the turbulence lasts well into effective matter dominated phase, any gravitational waves sourced via the turbulent mixing will be less strongly diluted. Consequently, the spectrum computed here represents a lower bound on the total signal. Further, the background will be strongly inhomogenous on small scales due to the bubble collisions and other preheating mechanisms. It is unclear exactly how efficient this preheating is and so how much it changes the effective equation of state, hence the last factor in \eq{eqn:energydensity} which may be significant. Thus, the suppression factor could again be reduced, leading to a stronger gravitational wave signal in the present epoch. These questions will most likely be settled via direct numerical simulations, and we will tackle these projects in future work.

\section*{Acknowledgments}

We thank Kenji Kadota for drawing our attention to an important error in an earlier draft of this paper. We would also like to thank Geraldine Servant for her comments on a draft of this paper.
WIP thanks Tufts University and Yale University for hospitality, and Jose Blanco-Pillado for encouraging this collaboration.
RE and JG are supported in part by the United States Department of Energy, grant DE-FG02-92ER-40704.
WIP and EDS are supported in part by
the Astrophysical Research Center for the Structure and Evolution of the Cosmos funded by the Korean Government (MOST) via the Korea Science and Engineering Foundation grant R01-2005-000-10404-0,
the Korea Research Foundation grant KRF-2005-201-C00006 funded by the Korean Government (MOEHRD),
and Brain Korea 21.
EAL is supported by the Institute of Strings, Cosmology and Astroparticle Physics at Columbia University, and would like to thank the Kavli Institute of Theoretical Physics (China) for its hospitality where some of this work was done.

\appendix

\section{Thermal tunnelling of the flaton}
\label{appendix:tunnelling}

The bubble nucleation rate is \cite{Linde:1980tt,Linde:1981zj,Linde:2005ht}
\begin{equation} \label{aag}
\Gamma \sim T^4 \exp{ \left[ - \frac{S_3(T)}{T} \right] }
\end{equation}
where
\begin{equation}
S_3(T) = S[\phi_\mathrm{c}(r),T] - S[0,T]
\end{equation}
\begin{equation}
S[\phi(r),T] = 4 \pi \int_0^\infty r^2 dr \left[ \frac{1}{2} \left( \frac{d\phi}{dr} \right)^2 + \fn{V}{\phi,T} \right]
\end{equation}
is the energy of a nucleated bubble, and $\phi_\mathrm{c}$ is the critical bubble (bounce solution) which minimizes the energy.
$\phi_\mathrm{c}$ satisfies
\begin{equation} \label{eqn:eom}
\frac{d^2\phi_\mathrm{c}}{dr^2} + \frac{2}{r} \frac{d\phi_\mathrm{c}}{dr} - \frac{\partial V}{\partial\phi} = 0
\end{equation}
with boundary conditions
\begin{eqnarray}
\fn{\frac{d\phi_\mathrm{c}}{dr}}{0} & = & 0
\\
\fn{\phi_\mathrm{c}}{\infty} & = & 0
\end{eqnarray}
Before percolation, the fraction of the universe consumed by bubbles is \cite{Hogan:1984hx}
\begin{equation} \label{fraction}
\fn{F}{t} \simeq \int_{-\infty}^t \d{t'} \fn{\Gamma}{t'} \frac{4\pi}{3} \fn{a}{t}^3 \left( \int_{t'}^t \frac{\d{t''}}{\fn{a}{t''}} \right)^3
\end{equation}
and the time of percolation, $t_\mathrm{p}$, is given by
\begin{equation} \label{tc1}
\fn{F}{t_\mathrm{p}} = 1
\end{equation}
During thermal inflation
\begin{equation}
\fn{a}{t} = e^{H_\mathrm{c} t}
\end{equation}
where $H_\mathrm{c}$ is given by \eq{hubble}.
Therefore \eq{fraction} becomes
\begin{equation} \label{FH}
\fn{F}{t} = \frac{4\pi}{3} \int_{-\infty}^t \d{t'} \fn{\Gamma}{t'} \frac{1}{H_\mathrm{c}^3} \left[ e^{ H_\mathrm{c} \left( t - t' \right) } - 1 \right]^3
\end{equation}
Assuming \footnote{See \ref{appendix:estimation} for numerical estimates.}
\begin{equation}
\frac{\beta}{H_\mathrm{c}} \equiv \frac{\dot\Gamma}{H_\mathrm{c} \Gamma} \gg 1
\end{equation}
then \eq{FH} is dominated by bubbles nucleated at times $t'$ well within a Hubble time of the final time
\begin{equation}
H_\mathrm{c} \left( t - t' \right) \ll 1
\end{equation}
and so
\begin{eqnarray}
\fn{F}{t} & \simeq & \frac{4\pi}{3} \int_{-\infty}^t \d{t'} \fn{\Gamma}{t'} \left( t - t' \right)^3
\\ \label{Fs}
& = & \frac{4\pi}{3} \int_0^\infty \d{s} s^3 \fn{\Gamma}{t-s}
\end{eqnarray}
where $s = t - t'$.
Taylor expanding $\ln\Gamma$ about the time of dominant bubble nucleation $t' = t_\mathrm{n}$ gives
\begin{equation}
\fn{\Gamma}{t-s} = \fn{\Gamma}{t_\mathrm{n}} \exp \left[ - \fn{\beta}{t_\mathrm{n}} \left( s - t + t_\mathrm{n} \right) + \ldots \right]
\end{equation}
where the higher order terms are small if \footnote{Numerical estimates give $|\dot\beta|/\beta^2 \sim 10^{-2} \textrm{ to } 10^{-3}$.}
\begin{equation} \label{linbeta}
|\dot\beta| \left( t' - t_\mathrm{n} \right)^2 \sim \frac{|\dot\beta|}{\beta^2} \ll 1
\end{equation}
Then \eq{Fs} becomes
\begin{eqnarray}
\fn{F}{t} & = & \frac{4\pi}{3} \int_0^\infty \d{s} s^3 \fn{\Gamma}{t_\mathrm{n}} \exp \left[ - \fn{\beta}{t_\mathrm{n}} \left( s - t + t_\mathrm{n} \right) \right]
\\
& = & \frac{4\pi}{3} \frac{\fn{\Gamma}{t_\mathrm{n}}}{\fn{\beta}{t_\mathrm{n}}^4} e^{ \fn{\beta}{t_\mathrm{n}} \left( t - t_\mathrm{n} \right) } \int_0^\infty \d{u} u^3 e^{-u}
\end{eqnarray}
where $u = \fn{\beta}{t_\mathrm{n}} s$.
Therefore
\begin{equation} \label{Ft}
\fn{F}{t} = \frac{8\pi \fn{\Gamma}{t_\mathrm{n}}}{\fn{\beta}{t_\mathrm{n}}^4} e^{ \fn{\beta}{t_\mathrm{n}} \left( t - t_\mathrm{n} \right) }
\end{equation}
with the dominant contribution coming from $u=3$.
Therefore the dominant bubbles at percolation are nucleated at
\begin{equation} \label{dominant}
t_\mathrm{n} = t_\mathrm{p} - \frac{3}{\fn{\beta}{t_\mathrm{n}}}
\end{equation}
and their size at percolation is
\begin{equation}
R = \frac{3}{\fn{\beta}{t_\mathrm{n}}}
\end{equation}
{}From \eqss{tc1}{Ft}{dominant}, $t_\mathrm{n}$ is given by
\begin{equation}
\fn{\Gamma}{t_\mathrm{n}} = \frac{\fn{\beta}{t_\mathrm{n}}^4}{8\pi e^3}
\end{equation}
Taking $t_{\mathrm{c}1} = t_\mathrm{n}$, reexpressing the time in terms of the temperature and using \eq{aag}, we define the $\beta$ parameter used in the body of the paper as
\begin{equation} \label{Tc1eq}
\frac{\beta}{H_\mathrm{c}}
\equiv \left. \frac{\dot{\Gamma}}{H \Gamma} \right|_{T_{\mathrm{c}1}}
\simeq \left[ T \frac{d}{dT} \left( \frac{S_3}{T} \right) \right]_{T_{\mathrm{c}1}}
\end{equation}
with $T_{\mathrm{c}1}$ determined by solving
\begin{equation}
\exp{ \left[ - \frac{\fn{S_3}{T_{\mathrm{c}1}}}{T_{\mathrm{c}1}} \right] } = \frac{1}{8\pi e^3} \left( \frac{\beta}{H_\mathrm{c}} \right)^4 \left( \frac{H_\mathrm{c}}{T_{\mathrm{c}1}} \right)^4
\end{equation}

\section{Flaton finite temperature effective potential}
\label{appendix:potential}

Before the phase transition, the flaton $\phi$ is strongly coupled to the thermal bath and its finite temperature effective potential is given by
\begin{equation} \label{Vfull}
V(\phi,T) = V^{(0)}(\phi) + V^{(1)}_T(\phi,T)
\end{equation}
\begin{equation}
V^{(0)}(\phi) = V_\mathrm{TI} - \frac{1}{2} m^2_\phi \phi^2 + \dots
\end{equation}
\begin{equation}
V^{(1)}_T(\phi,T) = T^4 \sum_p{ g_p J_p \left( \frac{m^2_p(\phi,T)}{T^2} \right) }
\end{equation}
where the sum is over all the particles in the thermal bath, $g_p$ is the number of degrees of freedom of the $p$th particle, $J_p = J_{\pm}$ for bosons and fermions respectively, and \cite{Dolan:1973qd}
\begin{eqnarray}
J_\pm(y^2) = \pm \frac{1}{2 \pi^2} \int^\infty_0 dx x^2 \ln{ \left( 1 \mp e^{-\sqrt{x^2 + y^2}} \right) }
\end{eqnarray}
The thermal effective mass-squareds of the non-MSSM quark superfields (which dominate the flaton's couplings to the thermal bath - see \ref{appendix:couplings}) are \cite{Comelli:1996vm,Donoghue:1983qx}
\begin{equation}
m_p^2(\phi,T) \simeq
\left\{
\begin{array}{ll}
m^2_\psi + \frac{1}{2} \lambda^2 \phi^2 + \left( \frac{1}{4} \lambda^2 + \frac{2}{3} g^2 \right) T^2 & \textrm{for bosons}
\\
\frac{1}{2} \lambda^2 \phi^2 + \frac{1}{6} g^2 T^2 & \textrm{for fermions}
\end{array}
\right.
\end{equation}
where we assume the boson masses $m_\psi \sim m_\mathrm{s}$ and the fermion masses are negligible.

\section{Flaton couplings to the thermal bath}
\label{appendix:couplings}

The flaton must have unsuppressed interactions with the thermal bath in order to be held at the origin during thermal inflation.
It is a gauge singlet so the interactions should be Yukawa couplings in the superpotential
\begin{equation} \label{coupling}
\sum_i \lambda_i \Phi \Psi_i \bar\Psi_i
\end{equation}
where $\Phi$ is the flaton superfield.
After thermal inflation, once the flaton has reached its vacuum value $\phi = \phi_0$, the superfields $\Psi_i$ and $\bar\Psi_i$ acquire masses
\begin{equation}
M_{\Psi_i} = \frac{\lambda_i \phi_0}{\sqrt{2}\,} \sim 10^{10} \textrm{ to } 10^{12} \GeV
\end{equation}
and so are not part of the MSSM.

The $\Psi_i$ and $\bar\Psi_i$ should form complete representations of SU(5) in order to preserve gauge coupling unification, and the size of the representations should be limited to at most two $\mathbf{10}$'s and two $\overline\mathbf{10}$'s in order to preserve perturbative gauge coupling unification \cite{Morrissey:2005uz}. The $\Psi_i$ and $\bar\Psi_i$ affect the renormalisation of the gauge couplings and their masses are proportional to $\phi$, therefore the gauge couplings will acquire a logarithmic dependence on $\phi$.
In particular, the strong gauge coupling during thermal inflation, $g_3(0)$, is related to the strong gauge coupling in the vacuum, $g_3(\phi_0) \simeq 1.2$, by
\begin{equation} \label{strong}
\frac{1}{g_3^2(0)} = \frac{1}{g_3^2(\phi_0)} + \frac{n_q}{16 \pi^2} \ln{ \left( \frac{\phi_0^2 + m_\mathrm{s}^2}{m_\mathrm{s}^2} \right) }
\end{equation}
where $3 n_q$ is the number of colored superfields in $\Psi$.
For example, $n_q = 1$ for $\Psi = \mathbf{5}$ and $n_q = 3$ for $\Psi = \mathbf{10}$.

At least some of the Yukawa couplings $\lambda_i$ should be strong enough to drive the flaton's mass-squared negative at low energies.
The renormalisation of the Yukawa couplings is given by
\begin{equation}
\frac{d \lambda_i}{d \ln{\Lambda}} = \frac{\lambda_i}{16 \pi^2} \left[ 3 \lambda_i^2 + \sum_{j \neq i} \lambda_j^2 - 4 \sum_a \fn{C_a}{R_i} g_a^2 \right]
\end{equation}
where $C_a(R_i)$ is the quadratic Casimir invariant for the superfield $i$ in representation $R_i$, for example $C_a = 4/3$ for a superfield in a $\mathbf{3}$ of SU(3).
Therefore the quark superfields will have the largest Yukawa couplings.
However, quarks in different multiplets could have different Yukawa couplings, so the number of quarks with strong Yukawa couplings $n_\lambda$ could be less than the total number of quarks $n_q$.
Strong quark Yukawa couplings will be pulled close to their renormalization group infrared fixed point
\begin{equation} \label{lambdaq}
\left( 3 n_\lambda + 2 \right) \lambda_\mathrm{FP}^2 = \frac{16}{3} \fn{g_3^2}{0}
\end{equation}
where $3 n_\lambda$ is the number of colored superfields with strong Yukawa couplings and $\lambda_\mathrm{FP}$ is the infrared fixed point of their Yukawa coupling.
Therefore, from \eq{strong},
\begin{equation}
\lambda_\mathrm{FP} = \frac{4 \fn{g_3}{\phi_0}}{ \sqrt{ \left( 9 n_\lambda + 6 \right) \left[ 1 + \frac{n_q}{16 \pi^2} \fn{g_3^2}{\phi_0} \ln{ \left( \frac{\phi_0^2 + m_\mathrm{s}^2}{m_\mathrm{s}^2} \right) } \right] }\, }
\end{equation}
Thus the strongest Yukawa couplings can be estimated and these will be the couplings that dominate the coupling of the flaton to the thermal bath.
Values of $\lambda_\mathrm{FP}$ for the plausible choices of representations are shown in Table~\ref{lambdatable}.
\begin{table}[ht]
\centering
\begin{tabular}{|c||c|c|c|c|c|c|c|c|} \hline
$\Psi$ & $\mathbf{5}$ & \multicolumn{2}{|c|}{$2 \times \mathbf{5}$} & $\mathbf{10}$ & $\mathbf{5} + \mathbf{10}$ & \multicolumn{2}{|c|}{$2 \times \mathbf{10}$} \\ \hline \hline
$n_q$ & 1 & \multicolumn{2}{|c|}{2} & 3 & 4 & \multicolumn{2}{|c|}{6} \\ \hline
$g_3(0)$ & 1.05 & \multicolumn{2}{|c|}{0.94} & 0.85 & 0.79 & \multicolumn{2}{|c|}{0.69} \\ \hline
$\Psi_\mathrm{eff}$ & $\mathbf{5}$ & $\mathbf{5}$ & $2 \times \mathbf{5}$ & $\mathbf{10}$ & $\mathbf{5} + \mathbf{10}$ & $\mathbf{10}$ & $2 \times \mathbf{10}$ \\ \hline \hline
$n_\lambda$ & 1 & 1 & 2 & 3 & 4 & 3 & 6 \\ \hline
$\lambda_\mathrm{FP}$ & 1.09 & 0.97 & 0.77 & 0.59 & 0.49 & 0.48 & 0.36 \\ \hline
\end{tabular}
\caption{ \label{lambdatable}
The infrared fixed point Yukawa couplings, $\lambda_\mathrm{FP}$, for $\phi_0 = 10^{11} \GeV$ and $m_\mathrm{s} = 10^3 \GeV$.
$\Psi_\mathrm{eff}$ are the representations with strong Yukawa couplings.
}
\end{table}

In the case of the baryogenesis scenario of Ref.~\cite{Jeong:2004hy}, the Affleck-Dine field $LH_u$ has an expectation value $l_0 \sim 10^9 \GeV$ at the end of thermal inflation, temporarily giving a large mass to some of the MSSM superfields.
Eq.~(\ref{strong}) then becomes
\begin{equation}
\frac{1}{g_3^2(0,l_0)} = \frac{1}{g_3^2(\phi_0,0)} + \frac{1}{16 \pi^2} \left[n_q \ln{ \left( \frac{\phi_0^2 + m_\mathrm{s}^2}{m_\mathrm{s}^2} \right) } - \sum_{q=u,c,t} \ln{ \left( \frac{\lambda_q^2 l_0^2 + m_\mathrm{s}^2}{m_\mathrm{s}^2} \right) } \right]
\end{equation}
with $\lambda_u \sim 10^{-5}$, $\lambda_c \sim 10^{-2}$ and $\lambda_t \sim 1$, and Table~\ref{lambdatable} is replaced by Table~\ref{lambdatable2}.
\begin{table}[ht]
\centering
\begin{tabular}{|c||c|c|c|c|c|c|c|c|} \hline
$\Psi$ & $\mathbf{5}$ & \multicolumn{2}{|c|}{$2 \times \mathbf{5}$} & $\mathbf{10}$ & $\mathbf{5} + \mathbf{10}$ & \multicolumn{2}{|c|}{$2 \times \mathbf{10}$} \\ \hline \hline
$n_q$ & 1 & \multicolumn{2}{|c|}{2} & 3 & 4 & \multicolumn{2}{|c|}{6} \\ \hline
$g_3(0)$ & 1.31 & \multicolumn{2}{|c|}{1.11} & 0.98 & 0.88 & \multicolumn{2}{|c|}{0.76} \\ \hline
$\Psi_\mathrm{eff}$ & $\mathbf{5}$ & $\mathbf{5}$ & $2 \times \mathbf{5}$ & $\mathbf{10}$ & $\mathbf{5} + \mathbf{10}$ & $\mathbf{10}$ & $2 \times \mathbf{10}$ \\ \hline \hline
$n_\lambda$ & 1 & 1 & 2 & 3 & 4 & 3 & 6 \\ \hline
$\lambda_\mathrm{FP}$ & 1.35 & 1.14 & 0.90 & 0.68 & 0.54 & 0.53 & 0.39 \\ \hline
\end{tabular}
\caption{ \label{lambdatable2}
The infrared fixed point Yukawa couplings, $\lambda_\mathrm{FP}$, for $\phi_0 = 10^{11} \GeV$, $m_\mathrm{s} = 10^3 \GeV$ and $l_0 = 10^9 \GeV$ in the case of the baryogenesis scenario of Ref.~\cite{Jeong:2004hy}.
}
\end{table}

\section{Estimation of $\beta / H_\mathrm{c}$}
\label{appendix:estimation}

We obtained the estimates for $\beta / H_\mathrm{c}$ given in Table~\ref{beta} by numerically integrating \eq{eqn:eom} and solving \eq{Tc1eq} for $T_{\mathrm{c}1}$, using the finite temperature effective potential given in \ref{appendix:potential} with the gauge couplings $g = \fn{g_3}{0}$ and Yukawa couplings $\lambda = \lambda_\mathrm{FP}$ given in Table~\ref{lambdatable}.
\begin{table}[h]
\centering
\begin{tabular}{|c|c||c|c|c|c|c|c|c|} \hline
\multicolumn{2}{|c||}{$\Psi$} & $\mathbf{5}$ & \multicolumn{2}{|c|}{$2 \times \mathbf{5}$} & $\mathbf{10}$ & $\mathbf{5} + \mathbf{10}$ & \multicolumn{2}{|c|}{$2 \times \mathbf{10}$} \\ \hline
\multicolumn{2}{|c||}{$\Psi_\mathrm{eff}$} & $\mathbf{5}$ & $\mathbf{5}$ & $2 \times \mathbf{5}$ & $\mathbf{10}$ & $\mathbf{5} + \mathbf{10}$ & $\mathbf{10}$ & $2 \times \mathbf{10}$ \\ \hline \hline
 & 0.5 & 900 & 1100 & 1600 & 3100 & 5600 & 6100 & 16000 \\ \cline{2-9}
\smash{$\displaystyle \frac{m_\psi}{m_\phi}$} & 1 & 1000 & 1300 & 1900 & 3800 & 7800 & 8300 & 23000 \\ \cline{2-9}
 & 2 & 1200 & 1600 & 2600 & 5700 & 12000 & 14000 & 39000 \\ \hline
\end{tabular}
\caption{\label{beta}
Estimates of $\beta/H_\mathrm{c}$ for $H_\mathrm{c} / T_{\mathrm{c}1} = 10^{-8}$. $m_\psi$ is the $\Psi$ boson mass and we set the $\Psi$ fermion mass equal to zero.
}
\end{table}

In the case of the baryogenesis scenario of Ref.~\cite{Jeong:2004hy}, using Table~\ref{lambdatable2}, Table~\ref{beta} is replaced by Table~\ref{beta2}.
\begin{table}[h]
\centering
\begin{tabular}{|c|c||c|c|c|c|c|c|c|} \hline
\multicolumn{2}{|c||}{$\Psi$} & $\mathbf{5}$ & \multicolumn{2}{|c|}{$2 \times \mathbf{5}$} & $\mathbf{10}$ & $\mathbf{5} + \mathbf{10}$ & \multicolumn{2}{|c|}{$2 \times \mathbf{10}$} \\ \hline
\multicolumn{2}{|c||}{$\Psi_\mathrm{eff}$} & $\mathbf{5}$ & $\mathbf{5}$ & $2 \times \mathbf{5}$ & $\mathbf{10}$ & $\mathbf{5} + \mathbf{10}$ & $\mathbf{10}$ & $2 \times \mathbf{10}$ \\ \hline \hline
 & 0.5 & 600 & 800 & 1100 & 2000 & 3700 & 4500 & 11000 \\ \cline{2-9}
\smash{$\displaystyle \frac{m_\psi}{m_\phi}$} & 1 & 700 & 900 & 1300 & 2500 & 4800 & 5900 & 17000 \\ \cline{2-9}
 & 2 & 800 & 1100 & 1600& 3400 & 7100 & 9100 & 26000\\ \hline
\end{tabular}
\caption{\label{beta2}
Estimates of $\beta/H_\mathrm{c}$ in the case of the baryogenesis scenario of Ref.~\cite{Jeong:2004hy}.
}
\end{table}

\end{document}